\documentclass[twocolumn,showpacs,aps,floatfix,superscriptaddress]{revtex4}
\usepackage{amsmath,amssymb,eucal,graphicx,bm}
\begin{document}
\title{First-Passage Exponents of Multiple Random Walks}
\author{E.~Ben-Naim}
\affiliation{Theoretical Division and Center for Nonlinear
Studies, Los Alamos National Laboratory, Los Alamos, New Mexico
87545, USA}
\author{P.~L.~Krapivsky}
\affiliation{Department of Physics,
Boston University, Boston, Massachusetts 02215, USA}
\begin{abstract}
We investigate first-passage statistics of an ensemble of $N$
noninteracting random walks on a line. Starting from a configuration
in which all particles are located in the positive half-line, we study
$S_n(t)$, the probability that the $n$th rightmost particle remains in
the positive half-line up to time $t$. This quantity decays
algebraically, $S_n (t)\sim t^{-\beta_n}$, in the long-time
limit. Interestingly, there is a family of nontrivial first-passage
exponents, $\beta_1<\beta_2<\cdots<\beta_{N-1}$; the only exception is
the two-particle case where $\beta_1=1/3$.  In the $N\to\infty$ limit,
however, the exponents attain a scaling form, $\beta_n(N)\to \beta(z)$
with $z=(n-N/2)/ \sqrt{N}$.  We also demonstrate that the smallest
exponent decays exponentially with $N$.  We deduce these results from
first-passage kinetics of a random walk in an $N$-dimensional cone and
confirm them using numerical simulations. Additionally, we investigate
the family of exponents that characterizes leadership statistics of
multiple random walks and find that in this case, the cone provides an
excellent approximation.
\end{abstract}
\maketitle

\section{Introduction}

Ensembles of ordinary random walks in one dimension
\cite{wf,ghw,hcb,rg} are used to model physical, chemical, and
biological processes ranging from wetting \cite{mef,hf} to the motion
of colloidal particles in narrow channels \cite{wbl,cdl} and
reaction-diffusion processes \cite{bpl}. In particular, first-passage
properties \cite{sr} of multiple random walks explain dynamics of
interacting spins \cite{dhp,snm}, lead changes in a voting process
\cite{wf,hn}, and lifetime of knots in polymer chains \cite{bdve}.

In these examples, first-passage properties are intertwined with the
ordering of the walkers
\cite{mef,hf,fg,bg,djg,kr,bjmkr,ck,dlb,yal}. Previous studies focused
on first-passage statistics of extreme particles; for example, the
probability that a single prey particle survives the predators to its
left \cite{bg,kr}. These problems involve a single first-passage
probability and hence, a single first-passage exponent \cite{sr}.

In this paper, we ask first-passage questions that concern the bulk
particles, not necessarily the extreme ones. We find that a family of
first-passage exponents characterizes the first-passage
kinetics. These exponents depend on two parameters: the particle order
and the total number of particles. Yet, when the number of particles
is very large, the exponents depend on a single scaling variable.
This scaling behavior is unusual. In equilibrium as well as in
non-equilibrium settings, one or two scaling exponents quantify a
scaling behavior \cite{gib,hes}. In the present case, however, the
exponents themselves obey scaling, and remarkably, there are scaling
laws for the scaling exponents.

\begin{figure}[t]
\includegraphics[width=0.42\textwidth]{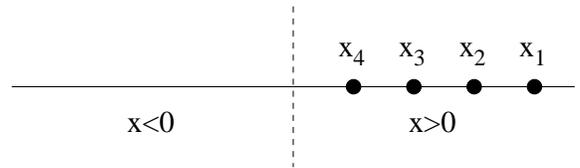}
\caption{Illustration of a four-particle system. Initially, all
particles are located in the positive half-line.  At time $t$, the
position of the rightmost particle is $x_1(t)$ and that of the
leftmost particle is $x_4(t)$.}
\label{fig-initial}
\end{figure}

We consider an array of $N$ identical particles that undergo simple
random walk on a one-dimensional line, and investigate two different
first-passage problems where the particle order plays an essential
role.  The first problem, discussed in Sections II-V, concerns the
probability that the $n$th rightmost particle hasn't crossed the
origin, if all particles start to the right of the origin (Figure
\ref{fig-initial}).  This quantity equals the likelihood that $N-n+1$
out of all random walks have yet to simultaneously reside in the
negative half-line. In the second problem, described in Section VI, we
identify the particle order with rank, such that the rightmost
particle is viewed as the leader, and similarly, the leftmost particle
as the laggard. We ask: what is the probability that the rank of the
initial leader does not fall below some specified threshold?

In both problems, we find a family of nontrivial first-passage
exponents. In both cases, the exponents depend on two variables: the
particle order and the total number of particles, $N$. Interestingly,
in the large-$N$ limit, the exponents become a function of a single
scaling variable. However, the similarities between the two problems
end here as the two scaling variables are fundamentally different and
moreover, the two scaling functions are dissimilar.

Our analysis relies on mapping the $N$ noninteracting random walks
onto a single compound random walk in $N$ dimensions. By combining
this mapping with exact and asymptotic properties for kinetics of
first-passage inside a cone \cite{bk}, we obtain approximate values
for the first-passage exponents.  The cone approximation is
straightforward to implement, yet it yields useful estimates for
exponents and in particular, this framework faithfully captures
typical and extremal properties of the first-passage exponents.

\section{The First-Passage Process}

Our system consists of $N$ identical particles. Each particle
undergoes a random walk on a one-dimensional lattice. At every time
step, one particle is selected at random, and it moves to the left,
$x\to x-1$, or to the right $x\to x+1$, with equal probabilities. Time
is augmented by the inverse number of particles after each such step,
$t\to t+1/N$. The particles always undergo independent random walks
and hence, they are noninteracting.

Let $x_n(t)$ be the position of the $n$th rightmost particle at time
$t$ (Figure \ref{fig-initial}).  We consider the initial configuration
where all particles are located in the positive half-line (Figure
\ref{fig-initial}), $x_n(0)>0$ for all $n$.  We stress that it is not
the initial order, but instead, the order at time $t$, that
sets the index $n$.

We are interested in $S_n(t)$, the probability that the $n$th
rightmost particle remains in the positive half-line until time
$t$. Hence, $S_n(t)$ is the likelihood that $x_n(\tau)>0$ for all
$0\leq \tau \leq t$.  In particular, $S_1(t)$ is the probability that
the rightmost particle has yet to cross the origin, while $S_N(t)$ is
the probability that the leftmost particle has not crossed the
origin. Clearly, the probabilities $S_n$ decrease monotonically with
$n$,
\begin{equation}
\label{monotonic}
S_1(t)\geq S_2(t)\geq \cdots \geq S_{N-1}(t)\geq S_N(t).
\end{equation}

The ``survival'' probabilities $S_n(t)$ generalize the classic
survival probability of a single one-dimensional random walk in the
presence of a trap. Indeed, when $N=1$, we have $S_1(t)\sim t^{-1/2}$
\cite{sr}. As usual, the survival probability $S_n$ immediately gives
the first-passage probability as $[-dS_n/dt]\times dt$ is the probability
that the $n$th rightmost particle crosses the origin for the first
time during the infinitesimal time interval $(t,t+dt)$.

\begin{figure}[t]
\includegraphics[width=0.45\textwidth]{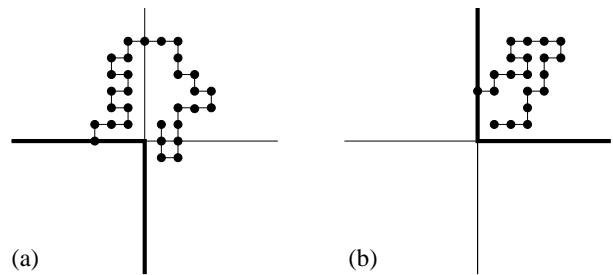}
\caption{The compound two-dimensional random walk.  The line indicates
  the random walk trajectory and the bullets show intermediate
  locations along this trajectory.  Thick lines mark the absorbing
  boundary.  (a) For $n=1$, the random walk is confined to the
  exterior of the third (negative) quadrant.  (b) For $n=2$, the
  random walk is confined to the interior of the first (positive)
  quadrant.}
\label{fig-rw2}
\end{figure}

The analytically solvable case of two particles yields valuable
insights into the general behavior.  When $N=2$, we map the two random
walks onto a single random walk in two dimensions. The position of the
two-dimensional walk is specified by the positions of the two
independent walks. At $t=0$, the two-dimensional walk is always inside
the first quadrant (Figure \ref{fig-rw2}).

The quantity $S_1(t)$ is the probability that the coordinates of the
two random walks have not become negative {\em simultaneously} up to
time $t$, or equivalently, the probability that the two-dimensional
walk remains in the exterior of the third quadrant (Figure
\ref{fig-rw2}a). To find the survival probability, we impose an
absorbing boundary condition along the edge of the third quadrant
(Figure \ref{fig-rw2}a). Then $S_1(t)$ is the probability that the
compound particle avoids this absorbing boundary up to time $t$.  The
region in which the random walk can move is a two-dimensional cone, or
equivalently, a wedge with opening angle $\alpha=3\pi/4$. (The opening
angle $\alpha$ is defined as the angle between the cone axis and the
cone surface, so it lies within the bounds $0\leq \alpha \leq \pi$.)
Thus, $S_1(t)$ equals the survival probability of a particle diffusing
inside a wedge with an absorbing surface. This survival probability
decays algebraically,
\begin{equation}
\label{wedge}
S(t)\sim t^{-\pi/4\alpha},
\end{equation}
in the long-time limit \cite{sr}. By substituting $\alpha=3\pi/4$ into
the general expression \eqref{wedge}, we find the intriguing behavior
\begin{equation}
\label{S12}
S_1(t)\sim t^{-1/3},
\end{equation}
as $t\to\infty$. As in \eqref{wedge}, this asymptotic behavior holds
regardless of the initial position, although the prefactor does depend
on the initial conditions.

Along the same lines, $S_2(t)$ is the probability that the positions
of the two random walks remain positive up to time $t$, or
alternatively, the probability that the two-dimensional walk remains
in the {\em interior} of the first quadrant (Figure
\ref{fig-rw2}b). Now, the boundary of the first quadrant is absorbing
(Figure \ref{fig-rw2}b), and the random walk is confined to a wedge
with opening angle $\alpha=\pi/4$. Using \eqref{wedge}, we again find
power-law behavior albeit with a larger exponent,
\begin{equation}
\label{S22}
S_2(t)\sim t^{-1}.
\end{equation}
As shown below, this particular behavior follows from an elementary
argument.

\section{A Family of Exponents}

The analytic results for a two-particle system suggest that generally,
the survival probabilities decay algebraically,
\begin{equation}
\label{Sn}
S_n(t)\sim t^{-\beta_n},
\end{equation}
in the long-time limit.  Moreover, we expect that the decay exponents
are distinct, and that $\beta_n\equiv \beta_n(N)$ increases
monotonically with $n$,
\begin{equation}
\label{betan}
\beta_1<\beta_2<\cdots<\beta_{N-1}<\beta_N,
\end{equation}
consistent with \eqref{monotonic}.

\begin{figure}[t]
\includegraphics[width=0.4\textwidth]{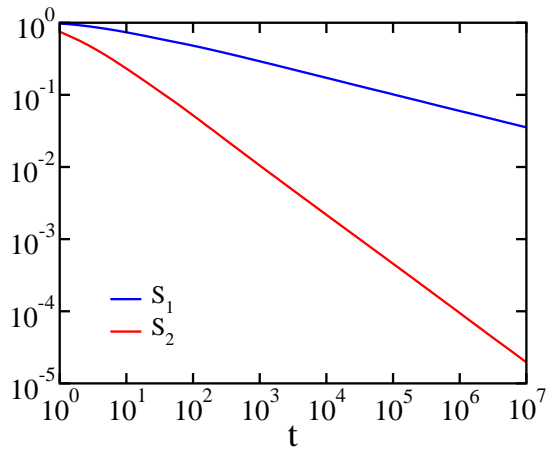}
\caption{The survival probabilities $S_1(t)$ and $S_2(t)$ for three
  particles. These results are from numerical simulations.}
\label{fig-s12}
\end{figure}

\begin{table}[ht]
\begin{tabular}{|c|l|l|l|l|l|l|}
\hline
N&$\beta_1$&$\beta_2$&$\beta_3$&$\beta_4$&$\beta_5$&$\beta_6$\\
\hline
$1$&1/2&&&&&\\
$2$&1/3&1&&&&\\
$3$&$0.228$&$0.68$&3/2&&&\\
$4$&$0.158$&$0.50$&$1.05$&2&&\\
$5$&$0.107$&$0.358$&$0.76$&$1.37$&5/2&\\
$6$&$0.071$&$0.261$&$0.584$&$1.05$&$1.7$&3
\\
\hline
\end{tabular}
\caption{The exponent $\beta_n$, obtained from numerical simulations for
$N\leq 6$.}
\end{table}

The largest exponent, $\beta_N$, is trivial. The probability $S_N(t)$
that the leftmost particle has yet to cross the origin equals the
probability that not a single particle crossed the origin. Since the
particles are independent, this probability equals the product of the
individual probabilities for each particle not to cross the origin,
$S_N\sim \left(t^{-1/2}\right)^N\sim t^{-N/2}$. Therefore, the largest
exponent is proportional to the total number of particles,
\begin{equation}
\label{largest}
\beta_N=\tfrac{1}{2}N\,.
\end{equation}

Our numerical simulations confirm that indeed, there are $N$ distinct
first-passage exponents.  Figure \ref{fig-s12} shows the survival
probabilities $S_1$ and $S_2$ for a three-particle system, and Table I
lists the exponents $\beta_n$ for $N\leq 6$. The simulations confirm
that there is a family of $N-1$ nontrivial exponents,
$\{\beta_1,\beta_2,\ldots,\beta_{N-1}\}$.

\begin{figure}[t]
\includegraphics[width=0.46\textwidth]{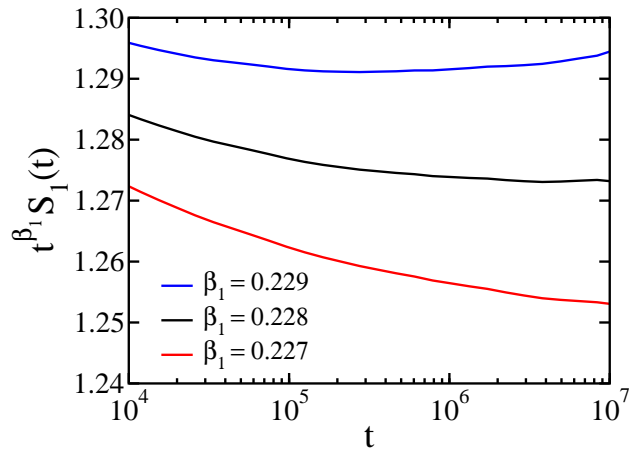}
\caption{The function $t^{\beta_1}\,S_1(t)$ versus $t$, using
  different values of $\beta_1$. The data is from numerical
  simulations of a three-particle system.}
\label{fig-slope}
\end{figure}

We performed extensive Monte Carlo simulations using the following
algorithm.  Initially, all $N$ particles are located at the same
position, $x=1$.  In each subsequent step one particle is selected at
random and it moves by one lattice site either to the left or to the
right with equal probabilities.  After each step, time is augmented by
the inverse number of particles, $t\to t+1/N$. Throughout this random
process, we keep track of the total number of particles in the region
$x<0$, and obtain the probability $S_n(t)$ from the fraction of
realizations in which this counter did not exceed $n$ until time
$t$. We computed the exponents $\beta_n$ by analyzing the local slope
$-d\ln S_n/d\ln t$ or alternatively, by seeking the value of $\beta_n$
for which the quantity $t^{\beta_n} S_n(t)$ achieves a plateau (Figure
\ref{fig-slope}).

We verified that the measured exponents $\beta_n$ are robust.  First,
we checked that they are not sensitive to the initial
positions. Second, we used parallel dynamics where particles move
synchronously, rather than sequentially, and found that the exponents
do not change.  The exponents reported in this study represent an
average over a large number of independent realizations. The total
number of realizations varies from $10^6$ for slow first-passage
processes ($\beta\ll 1$) to as high as $10^{12}$ for fast processes
($\beta\gg 1$). We note that it is difficult to measure tiny exponents
$\beta\ll 1$ because the simulations must run to very large times, and
furthermore, it is difficult to obtain large exponents $\beta\gg 1$,
because now, we need an enormous number of realizations.

As a side note, the behavior \eqref{Sn}-\eqref{betan} has a natural
interpretation in the context of single-file diffusion where the
random walks interact by hard core exclusion
\cite{teh,dgl,ap,vkk,tw,ra,krb}. Under the standard transformation
where the identities of two particles are exchanged whenever their
trajectories cross, the noninteracting particle system is equivalent
to an interacting particle system.  In the context of this exclusion
process, the quantity $S_n(t)$ is the probability that the $n$th out
of $N$ particles avoids the negative half-line up to time $t$.

\section{Cone Approximation}

In general, we map the $N$ one-dimensional random walks onto a single
random walk in $N$ dimensions. To find the survival probability
$S_n(t)$, we require that the random walk remains inside the region in
which the number of negative coordinates is smaller than $n$ up to the
$t$.  This region occupies a fraction $V_n$ of space given by
\begin{equation}
\label{Vn}
V_n = 2^{-N}\sum_{m=n}^N \binom{N}{m}.
\end{equation}
For example, when $N=2$ we have $(V_1,V_2)=(3/4,1/4)$ as in
Fig.~\ref{fig-rw2}, while $(V_1,V_2,V_3)=(7/8,1/2,1/8)$ for $N=3$.

In two dimensions, the random walk is confined to two-dimensional
cones of opening angles $\alpha=3\pi/4$ and $\alpha=\pi/4$ when $n=1$
and $n=2$, respectively. In the cone approximation, we replace the
region of space that confines the walk with an {\em unbounded}
circular cone in $N$ dimensions that occupies the same fraction of
space $V_n$. An $N$-dimensional cone with opening angle $\alpha$
occupies the fraction of space
\begin{equation}
\label{Valpha}
V(\alpha)=\frac{\int_0^\alpha d\theta\,(\sin\theta)^{N-2}} {\int_0^\pi
d\theta\,(\sin\theta)^{N-2}}.
\end{equation}
Using a spherical coordinate system, this expression follows from
$d\Omega\propto(\sin\theta)^{N-2}d\theta$ where $\Omega$ is the solid
angle and $\theta$ is the polar angle.  We now choose the cone opening
angle $\alpha\equiv \alpha(n,N)$ as follows
\begin{equation}
\label{equal}
V(\alpha)=V_n.
\end{equation}

\begin{table}[t]
\begin{tabular}{|c|l|l|l|l|l|l|}
\hline
N&$\beta_1^{\rm cone}$&$\beta_2^{\rm cone}$&$\beta_3^{\rm cone}$&$
\beta_4^{\rm cone}$&$\beta_5^{\rm cone}$&$\beta_6^{\rm cone}$\\
\hline
$2$&1/3&1&&&&\\
$3$&$0.216785$&$1/2$&$1.407279$&&&\\
$4$&$0.139844$&$0.337987$&$0.739671$&$1.787705$&&\\
$5$&$0.0893854$&$0.237106$&$1/2$&$0.998423$&$2.151625$&\\
$6$&$0.0564058$&$0.166833$&$0.359218$&$0.684043$&$1.265701$&$2.503786$
\\
\hline
\end{tabular}
\caption{The exponents $\beta_n^{\rm cone}$, obtained using the cone
  approximation, that is, Eqs.~\eqref{Vn}--\eqref{cone}.}
\end{table}

The survival probability of a particle that diffuses inside an
unbounded $N$-dimensional cone decays algebraically, $S\sim
t^{-\beta}$, in the long-time limit \cite{sr,rdd,bs}. The exponent
$\beta\equiv \beta(\alpha)$ decreases as the opening angle $\alpha$
increases. In two dimensions, $\beta=\pi/4\alpha$ as in \eqref{wedge},
and in four dimensions $\beta=(\pi-\alpha)/2\alpha$ \cite{bk}.
Generally, the decay exponent $\beta$ is the smallest root of a
transcendental equation involving either $P_{2\beta+\delta}^\delta$ or
$Q_{2\beta+\delta}^\delta$, that is, one of the two associated
Legendre functions \cite{NIST} of degree $2\beta+\delta$ and order
$\delta=\frac{N-3}{2}$ \cite{bk},
\begin{equation}
\label{cone}
\begin{split}
P_{2\beta+\delta}^\delta(\cos\alpha) &= 0\qquad N\ {\rm odd},\\
Q_{2\beta+\delta}^\delta(\cos\alpha) &= 0\qquad N\ {\rm even}.
\end{split}
\end{equation}
In three dimensions, the transcendental equation involves the Legendre
function, $P_{2\beta}(\cos\alpha)=0$. In all dimensions, a cone with
$\alpha=\pi/2$ is simply a plane, and hence $\beta(\pi/2)=1/2$.

By construction, the cone approximation is exact in two
dimensions. For three particles, the cone approximation provides good
estimates for the smallest and the largest exponents. When $n=1$, the
cone specified by \eqref{equal} has opening angle
$\alpha_1=\pi-0.722734$ , and equation \eqref{cone} gives the
first-passage exponent $\beta_1^{\rm cone}=0.216785$ where the
superscript indicates outcome of the cone approximation.  This value
represents a very good approximation, as the simulations yield
$\beta_1=0.228$. For the largest exponent ($n=3$) the opening angle is
$\alpha_3=0.722734$ and the consequent value $\beta_3^{\rm
  cone}=1.407279$ is close to the exact value $\beta_3=3/2$. When
$n=2$, there is a larger discrepancy: $\alpha_2=\pi/2$ leads to the
approximate value $\beta_2^{\rm cone}=1/2$, while the simulations
yield $\beta_2=0.68$.

Table II lists the outcome of the cone approximation for $2\leq
N\leq6$.  Comparing with the respective values in Tables I, we see
that quantitatively, the cone approximation deteriorates as $N$
grows. Nevertheless, the cone approximation faithfully captures {\em
  all} qualitative features of the first-passage exponents, as shown
below.

\section{Scaling and Extremal Properties}

We are especially interested in the behavior when the number of
particles is large, $N\gg 1$.  Let us first evaluate the fraction
$V_n$ in the limit $N\to\infty$.  Using the Stirling formula $N!\simeq
\sqrt{2\pi N}\, N^N\,e^{-N}$ we write the leading behavior of the
binomial term in \eqref{Vn},
\begin{equation*}
2^{-N}\binom{N}{m}\simeq \sqrt{2/\pi N}\,e^{-2(m-N/2)^2/N}.
\end{equation*}
We now take the continuum limit and convert the sum on the right hand
side of \eqref{Vn} into an integral. The quantity $V_n$ becomes a
function of a single scaling variable
\begin{equation}
\label{Vn-scaling}
V_n(N)\to \frac{1}{2}\,{\rm erfc}
\left(z\sqrt{2}\right)\quad {\rm with}\quad z=\frac{n-N/2}{\sqrt{N}},
\end{equation}
where ${\rm erfc}(\xi)=(2/\sqrt{\pi})\int_\xi^\infty e^{-u^2}du$ is
the complementary error function.  Equation \eqref{Vn-scaling} is
valid in the limit \hbox{$n,N\to\infty$} with the scaling variable
$(n-N/2)/\sqrt{N}$ finite.

Similarly, we evaluate the leading large-$N$ behavior of $V(\alpha)$
given in \eqref{Valpha}.  The dominant contribution to the integral
comes from a narrow region of order $1/\sqrt{N}$ in the vicinity of
$\alpha=\pi/2$ where the integrand is Gaussian,
\begin{equation*}
(\sin\theta)^{N-2} \simeq e^{-N(\pi/2-\theta)^2/2}.
\end{equation*}
Using the leading asymptotic behavior of the denominator,
\hbox{$\int_{-\infty}^\infty
  \exp\big[\!-\!N(\pi/2-\theta)^2/2\big]d\theta\to \sqrt{2\pi/N}$}, we
find that the fraction $V(\alpha)$ has the scaling form
\begin{equation}
\label{Valpha-scaling}
V(\alpha,N)\to \frac{1}{2}\,{\rm erfc}\left(\frac{y}{\sqrt{2}}\right)
\quad {\rm with}\quad
y=(\cos\alpha)\sqrt{N}.
\end{equation}
In writing this equation, we used the fact
\hbox{$\cos\alpha\simeq\pi/2-\alpha$} when $\pi/2-\alpha\to 0$.
Equation \eqref{Valpha-scaling} is valid in the limit $\pi/2-\alpha\to
0$, $N\to\infty$, with the scaling variable $(\cos\alpha)\sqrt{N}$
finite. Meanwhile, asymptotic analysis of \eqref{cone} shows that in
the limit $N\to\infty$, $\pi/2-\alpha\to 0$ with the scaling variable
$(\cos\alpha)\sqrt{N}$ finite, the exponent $\beta(\alpha)$ becomes a
function of the scaling variable $y$ alone,
\begin{equation}
\label{betaalpha-scaling}
\beta(\alpha,N)\to \beta(y)\quad{\rm with}\quad y=(\cos\alpha)\sqrt{N}.
\end{equation}
The dependence of the exponent $\beta$ on the scaling variable $y$ is
specified through the transcendental equation $D_{2\beta}(y)=0$, where
$D_\nu$ is the parabolic cylinder function \cite{NIST} of order $\nu$,
and the acceptable root is the smallest one \cite{bk}.

\begin{figure}[t]
\includegraphics[width=0.45\textwidth]{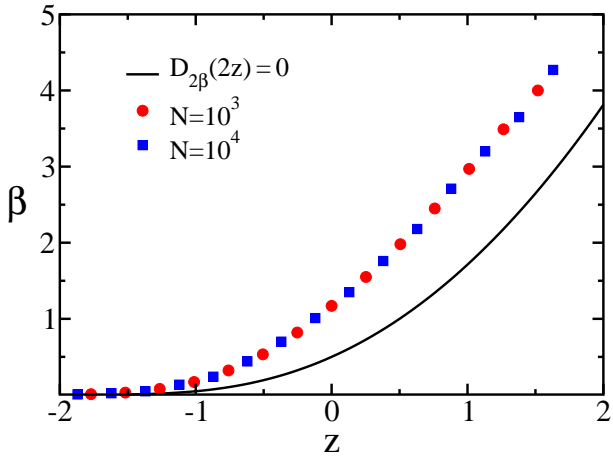}
\caption{The exponent $\beta$ versus the scaling variable $z$. The
  Monte Carlo results are from simulations with $N=10^3$ (bullets) or
  $N=10^4$ (squares) particles.  The solid line is the scaling
  function \eqref{scaling-function} obtained using the cone
  approximation.}
\label{fig-beta}
\end{figure}

Comparing equation \eqref{Vn-scaling} with equation
\eqref{Valpha-scaling}, we conclude that the first-passage exponent
depends on a single scaling variable,
\begin{equation}
\label{beta-scaling}
\beta_n(N) \to \beta(z) \quad {\rm with}\quad z=\frac{n-N/2}{\sqrt{N}},
\end{equation}
in the large-$N$ limit. Using $y=2z$, the exponent $\beta$ and the
scaling variable $z$ are related by the transcendental equation
\begin{equation}
\label{scaling-function}
D_{2\beta}(2z)=0.
\end{equation}
In particular, $\beta(z=0)=1/2$ (Figure \ref{fig-beta}).

The scaling behavior \eqref{beta-scaling}--\eqref{scaling-function}
implies that the exponents $\beta_n$ are of order one only in a window
of size $\sqrt{N}$ centered on the mid-point $n=N/2$.  Otherwise, the
exponents are exponentially small when $N/2-n\gg \sqrt{N}$, or
algebraically large when $n-N/2\gg \sqrt{N}$.  The limiting behaviors
of the scaling function captures these extremal properties,
\begin{equation}
\label{scaling-limits}
\beta^{\rm cone}(z)\simeq
\begin{cases}
\sqrt{z^2/2\pi}\,\exp\!\left(-2z^2\right)&z\to-\infty\\
z^2/2 & z\to\infty.
\end{cases}
\end{equation}
Both of these limiting behaviors follow from \eqref{scaling-function},
see \cite{bk}. The algebraic behavior in the limit $z\to\infty$
implies $\beta_N\sim N$, in agreement with \eqref{largest}.

Results of massive numerical simulations with a large number of
particles confirm (Fig.~\ref{fig-beta}) that the first-passage
exponents adhere to the scaling form \eqref{beta-scaling}.  Moreover,
the shape of the scaling function is qualitatively similar to the
shape of the scaling function \eqref{scaling-function}.

\begin{figure}[t]
\includegraphics[width=0.47\textwidth]{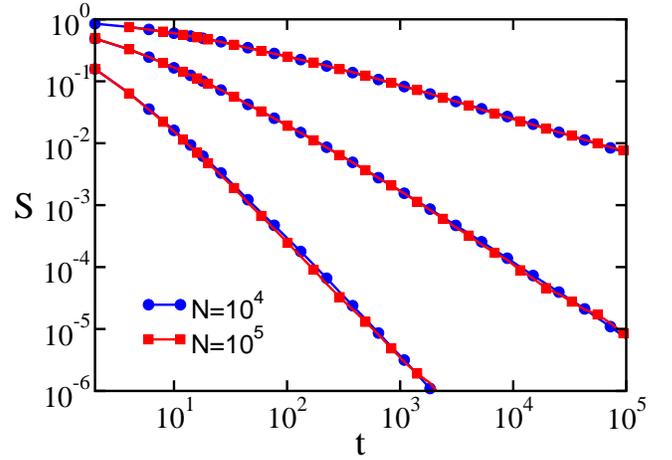}
\caption{The survival probability $S(z,t)$ versus time $t$.  Shown are
  results of simulations with different number of particles, $N=10^4$
  (bullets) and $N=10^5$ (squares), at three different values of the
  scaling variable, $z=-1/2$ (upper curves), $z=0$ (intermediate
  curves), and $z=1/2$ (lower curves).}
\label{fig-st}
\end{figure}

Additionally, the simulations reveal (Fig.~\ref{fig-st}) that not only
the scaling exponents depend on a single scaling variable, but the
entire survival probability is also function of the scaling variable
$z$. That is,
\begin{equation}
\label{Sn-scaling}
S_n(t) \to S(z,t) \quad {\rm with}\quad z=\frac{n-N/2}{\sqrt{N}}
\end{equation}
when $N\gg 1$. We confirmed this behavior numerically using a very
large number of particles (Figure \ref{fig-st}).  The scaling behavior
\eqref{Sn-scaling} is notable because it applies at all times: it
holds at short times, at intermediate times, and at long times.

Of special interest is the survival probability of the median particle
which becomes completely independent of the total number of particles
when $N\to\infty$. There is an interesting limiting value that
characterizes the survival probability of the median particle,
\begin{equation}
\beta_{\rm mid}=\lim_{N\to\infty}\beta_{N/2}.
\end{equation}
Numerically, $\beta_{\rm mid}=1.16$, while the cone approximation
gives $\beta_{\rm mid}=1/2$. This universal behavior and the scaling
form \eqref{beta-scaling} show that there is a narrow
``scaling-window'' of width $\Delta n$ with $\Delta n/N\sim N^{-1/2}$
centered on the median particle. Hence, only the relatively small
number of particles residing inside this region influence the median
particle. This behavior is in line the narrow range of cone opening
angles, $\Delta \alpha\sim N^{-1/2}$, where the first-passage
exponents change rapidly according to \eqref{betaalpha-scaling}.

\begin{figure}[t]
\includegraphics[width=0.45\textwidth]{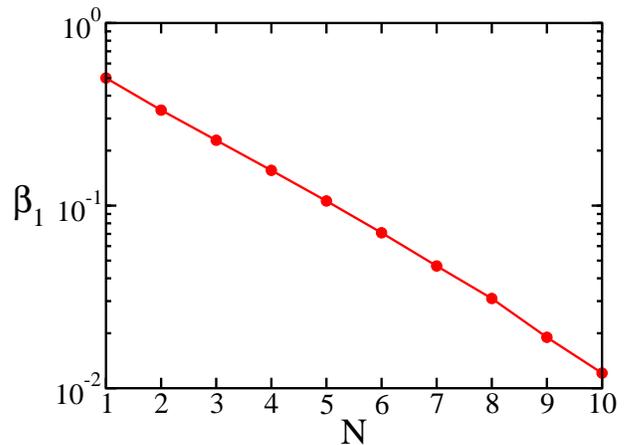}
\caption{The smallest exponent, $\beta_1$, versus the total number of
  particle $N$. The results are from Monte Carlo simulations.}
\label{fig-beta1}
\end{figure}

The smallest exponent $\beta_1$ underlying the decay of $S_1(t)$ is
especially intriguing because $S_1(t)$ has a simple interpretation ---
it is the probability that the $N$-dimensional walk remains in the
exterior of the generalized $N$-dimensional ``quadrant'' (see Figure
\ref{fig-rw2}), a region that occupies a fraction $V_1=1-2^{-N}$ of
space.  Table III shows that the cone approximation yields useful
estimates for $\beta_1$ when $N$ is small. The cone approximation is
also useful for understanding the large-$N$ behavior.  Substituting
\hbox{$V_1=1-2^{-N}$} into \eqref{equal} and
$[\sin(\pi-\theta)]^N\simeq (\pi-\theta)^N$ into \eqref{Valpha}, we
readily find that the opening angle approaches a constant,
\begin{equation}
\alpha\to\pi-\frac{1}{2},
\end{equation}
in the limit $N\to\infty$. In a cone of fixed opening angle
$\alpha>\pi/2$, the survival exponent $\beta$ shrinks exponentially
with dimension, $\beta\sim [\sin(\pi-\alpha)]^N$ in the large-$N$
limit \cite{bk}.  Thus, we conclude that the smallest exponent decays
exponentially with the total number of particles,
\begin{equation}
\label{beta1}
\beta_1\sim e^{-C\,N},
\end{equation}
with \hbox{$C^{\rm cone}=-\ln (\sin \tfrac{1}{2})=0.735167$}. The
simulation results are consistent with this exponential decay
(Fig.~\ref{fig-beta1}).

\begin{table}[t]
\begin{tabular}{|c|l|l|}
\hline
N&$\beta_1^{\rm cone}$&$\beta_1$\\
\hline
$2$&$1/2$&$1/2$\\
$3$&$0.216785$&$0.228$\\
$4$&$0.139844$&$0.156$\\
$5$&$0.0893854$&$0.106$\\
$6$&$0.0564058$&$0.0710$\\
$7$&$0.0350414$&$0.0467$\\
$8$&$0.0213967$&$0.0310$\\
$9$&$0.01283556$&$0.0189$\\
$10$&$0.00756816$&$0.0123$\\
\hline
\end{tabular}
\caption{The smallest exponent as a function of $N$.  Listed are the
  outcome of the cone approximation, $\beta_1^{\rm cone}$, and the
  results of numerical simulations, $\beta_1$.}
\end{table}

The exponential decay \eqref{beta1} and the algebraic growth
\eqref{betan} imply the following limiting behaviors of $\beta$ in the
scaling regime,
\begin{equation}
\label{scaling-limits-1}
\beta(z)\sim
\begin{cases}
\exp\left(-{\rm const.}\times z^2\right)&z\to-\infty,\\
z^2 & z\to\infty.
\end{cases}
\end{equation}
These behaviors are fully consistent with
\eqref{scaling-limits}. Therefore, the cone approximation captures all
qualitative features of the first-passage exponents including typical
behavior that includes the form of the scaling variable and the shape
of the scaling function, as well as extremal behavior comprising of
exponential decay of the smallest exponent and algebraic growth of the
largest exponent \cite{largest}.

\section{Leadership Statistics}

The cone approximation is useful in other contexts.  We now apply this
approach to understand leadership statistics of multiple random walks.

We study the very same system of $N$ independent random walks in one
dimension. As shown in Figure \ref{fig-initial}, we rank the particles
by position with $n=1$ being the rightmost particle and $n=N$ being
the leftmost particle.  As if the random walks participate in a
competition \cite{kr,bjmkr,bh,bmr,abk}, we view the rightmost particle as
leader and the leftmost particle as laggard. We now investigate
$P_n(t)$, the probability that the rank of the particle initially in
the lead, does not fall below $n$ up to time $t$.  Thus, $P_1(t)$ is
the probability that the original leader never loses the lead, while
$P_{N-1}(t)$ is the probability that the original leader does not
become the laggard.

If $N=2$, we have $P_1\sim t^{-1/2}$. For three particles, $P_1(t)\sim
t^{-3/4}$ and $P_2\sim t^{-3/8}$ \cite{bjmkr}. In this case, the
compound random walk in two dimensions is confined to wedges formed by
two intersecting planes.  These wedges occupy fractions $V_1=1/3$ and
$V_2=2/3$ of space, and the decay exponents are given by
$\gamma=\pi/4\alpha$ or equivalently, $\gamma=1/4V$.  Based on the
results for two- and three-particle systems, we expect that the
probabilities $P_n$ decay algebraically,
\begin{equation}
P_n(t)\sim t^{-\gamma_n},
\end{equation}
in the long-time limit. Once again, there is a family of first-passage
exponents,
\begin{equation}
\gamma_1>\gamma_2>\cdots>\gamma_{N-1}.
\end{equation}
The largest exponent $\gamma_1$ characterize the probability that the
initial leader maintains the lead, and the smallest exponent
$\gamma_{N-1}$ characterizes the probability that the original leader
never turns into the laggard.

In this case, the compound $N$-dimensional walk moves in a region that
occupies a fraction 
\begin{equation}
\label{Vn-gen}
V_n=\frac{n}{N} 
\end{equation}
of space. For example, to see that $V_1=1/N$ we note that the
total $N$-dimensional space is divided into $N$ equivalent regions in
which the identity of the leader is the same.  It is straightforward
to generalize this result to all $n$.  We have seen that when there
are three particles, the boundary of the allowed region forms a {\em
two-dimensional} cone. We therefore replace the allowed region with an
$N-1$ dimensional cone occupying the same fraction of space 
\eqref{Vn-gen} as the allowed region. The opening angle of this cone,
$\alpha$, satisfies 
\begin{equation}
\label{alpha-lead}
\frac{\int_0^\alpha d\theta\,(\sin\theta)^{N-3}} {\int_0^\pi
d\theta\,(\sin\theta)^{N-3}}=\frac{n}{N}.
\end{equation}
Given $n$ and $N$, we first determine the opening angle $\alpha$ by
solving \eqref{alpha-lead}, and then compute the exponent $\gamma$ as
the smallest root of a transcendental equation involving one of the
two associated Legendre functions with degree $2\gamma+\mu$ and order
$\mu=\frac{N-4}{2}$,
\begin{equation}
\label{cone1}
\begin{split}
P_{2\gamma+\mu}^{\mu}(\cos\alpha) &= 0\qquad N\ {\rm even},\\
Q_{2\gamma+\mu}^{\mu}(\cos\alpha) &= 0\qquad N\ {\rm odd}.
\end{split}
\end{equation}
By construction, the cone approximation is exact for three particles.
The cone approximation values, listed in Table IV, provide very good
estimates, given the Monte Carlo simulation values listed in Table
V. The numerical simulations are implemented using the algorithm
described in Section III, except that at time $t=0$, the random walks
occupy $N$ consecutive lattice sites.

\begin{table}[t]
\begin{tabular}{|c|l|l|l|l|l|}
\hline
N&$\gamma_1^{\rm cone}$&$\gamma_2^{\rm cone}$&$\gamma_3^{\rm cone}$&
$\gamma_4^{\rm cone}$&$\gamma_5^{\rm cone}$\\
\hline
$3$&$3/4$&$3/8$&&&\\
$4$&$0.888644$&$1/2$&$0.300754$&&\\
$5$&$0.986694$&$0.612148$&$0.408397$&$0.253371$&\\
$6$&$1.062297$&$0.701508$&$1/2$&$0.351915$&$0.220490$\\
\hline
\end{tabular}
\caption{The exponents $\gamma_n^{\rm cone}$, obtained using a cone in
$N-1$ dimensions with fraction of space $V_n=n/N$.}
\end{table}

\begin{table}[t]
\begin{tabular}{|c|l|l|l|l|l|}
\hline
N&$\gamma_1$&$\gamma_2$&$\gamma_3$&$\gamma_4$&$\gamma_5$\\
\hline
$2$&1/2&&&&\\
$3$&$3/4$&$3/8$&&&\\
$4$&$0.913$&$0.556$&$0.306$&&\\
$5$&$1.02$&$0.676$&$0.454$&$0.265$&\\
$6$&$1.11$&$0.769$&$0.557$&$0.391$&$0.234$\\
\hline
\end{tabular}
\caption{The exponent $\gamma_n$ from numerical simulations.}
\end{table}

\begin{figure}[t]
\includegraphics[width=0.45\textwidth]{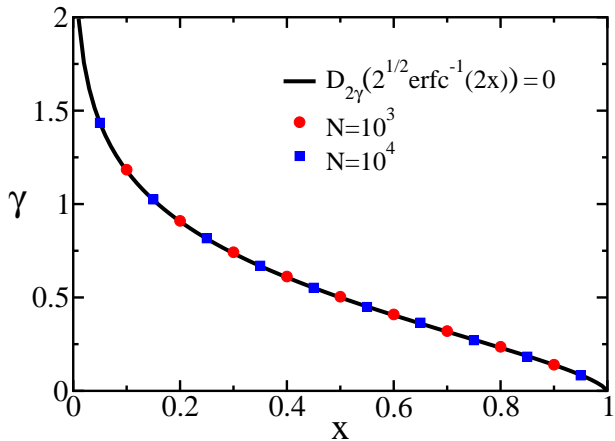}
\caption{The exponent $\gamma$ versus the scaling variable $x$. The
line shows the prediction of the cone approximation given in equation
\eqref{gamma-scaling}, as well as results of numerical simulations
with $N=10^3$ (bullets) and $N=10^4$ particles (squares).}
\label{fig-gamma}
\end{figure}

From \eqref{alpha-lead}, we expect that the exponent $\gamma$ depends
on a single scaling variable,
\begin{equation}
\label{gammax}
\gamma_n(N)\to \gamma(x)\quad{\rm with}\quad x=\frac{n}{N},
\end{equation}
in the limit $N\to\infty$.  When $N$ is very large, we can replace the
left-hand side in \eqref{alpha-lead} with the right-hand side in
\eqref{Valpha}.  Substituting \eqref{Vn-gen} into
\eqref{Valpha-scaling} shows that
\begin{equation}
\label{y-newdef}
y=\sqrt{2}\,{\rm erfc}^{-1}(2x).
\end{equation}
Here ${\rm erfc}^{-1}(x)$ is the inverse complementary error function,
(see below equation \eqref{Vn-scaling}).  Using the scaling behavior
\eqref{scaling-function}, we find that $\gamma$ and $x$ are related by
the transcendental equation (Figure \ref{fig-gamma})
\begin{equation}
\label{gamma-scaling}
D_{2\gamma}\left(\sqrt{2}\,{\rm erfc}^{-1}(2x)\right)=0,
\end{equation}
where $D_\nu$ is the parabolic cylinder function. We note that
$\gamma(0)=1/2$.  Given the definition of $x$, the scaling behavior in
the leadership problem is completely different than the scaling
behavior in the origin-crossing problem.

\begin{table}[t]
\begin{tabular}{|c|l|l|}
\hline
N&$\gamma_{N-1}^{\rm cone}$&$\gamma_{N-1}$\\
\hline
$2$&$1/2$&$1/2$\\
$3$&$3/8$&$3/8$\\
$4$&$0.300754$&$0.306$\\
$5$&$0.253371$&$0.265$\\
$6$&$0.220490$&$0.234$\\
$7$&$0.196216$&$0.212$\\
$8$&$0.177469$&$0.190$\\
$9$&$0.162496$&$0.178$\\
$10$&$0.150221$&$0.165$\\
\hline
\end{tabular}
\caption{The smallest exponent as a function of $N$.  Listed are
estimates from the cone approximation, $\gamma_{N-1}^{\rm cone}$, and
results of numerical simulations, $\gamma_{N-1}$.}
\end{table}

\begin{table}[t]
\begin{tabular}{|c|l|l|}
\hline
N&$\gamma_1^{\rm cone}$&$\gamma_1$\\
\hline
$3$&$3/4$&$3/4$\\
$4$&$0.888644$&$0.91$\\
$5$&$0.986694$&$1.02$\\
$6$&$1.062297$&$1.11$\\
$7$&$1.123652$&$1.19$\\
$8$&$1.175189$&$1.27$\\
$9$&$1.219569$&$1.33$\\
$10$&$1.258510$&$1.37$\\
\hline
\end{tabular}
\caption{The largest exponent as a function of $N$.  Listed are the
estimates from the cone approximation, $\gamma_1^{\rm cone}$, and results
of numerical simulations, $\gamma_1$.}
\end{table}

Using the asymptotic behaviors
\begin{equation*}
{\rm erfc}(\xi)\simeq
\begin{cases}
2-\big(\sqrt{1/\pi \xi^2}\,\big)\exp(-\xi^2)& \xi\to -\infty, \\
\big(\sqrt{1/\pi \xi^2}\,\big)\,\exp(-\xi^2) & \xi\to\infty,
\end{cases}
\end{equation*}
we deduce that the variable $y$ defined in \eqref{y-newdef} has
the asymptotic behaviors
\begin{equation*}
y(x)\simeq
\begin{cases}
\sqrt{2\ln \frac{1}{2x}} & x\to 0, \\
-\sqrt{2\ln \frac{1}{2(1-x)}} & x\to 1.
\end{cases}
\end{equation*}
Substituting these expressions into the asymptotic behavior of
the scaling function in \eqref{betaalpha-scaling}, we find the
limiting behaviors (see also \cite{bk})
\begin{equation}
\label{gamma-scaling-limits}
\gamma(x)\simeq
\begin{cases}
\frac{1}{4}\ln \frac{1}{2x}&x\to 0,\\
(1-x)\ln \frac{1}{2(1-x)} & x\to 1.
\end{cases}
\end{equation}
The exponent $\gamma$ decreases monotonically with
$x$: it weakly diverges when $x\to 0$ and
it vanishes as $x\to 1$.  By substituting $x=1/N$ and $1-x=1/N$,
respectively, into the appropriate expressions in
\eqref{gamma-scaling-limits}, we find the leading large-$N$
behaviors
\begin{equation}
\label{gamman-limits}
\gamma_n\simeq
\begin{cases}
\frac{1}{4}\ln N&n=1,\\
\frac{1}{N}\ln N & n={N-1}.
\end{cases}
\end{equation}
Both expressions match estimates based on heuristic arguments
\cite{kr,bjmkr}.  The smallest and the largest exponents, listed
respectively in Tables VI-VII, show that the cone yields an excellent
approximation. Yet, the quality of the approximation declines ever so
slightly as $N$ increases.

Monte Carlo simulations with a large number of particles confirm the
scaling behavior \eqref{gammax}. Moreover, we numerically verified
that the entire survival probability becomes a universal function of
the scaling variable $x$, in analogy with \eqref{Sn-scaling}, that is
$P_n(N,t)\to P(x,t)$ with $x=n/N$, as $N\to\infty$. Interestingly, the
numerical results strongly suggest that the scaling function specified
in \eqref{gamma-scaling} is asymptotically {\em exact} (Figure
\ref{fig-gamma}).  Finally, the exponent $\gamma_{N/2}$, that
characterizes the probability that the original leader always ranks
higher than median has a simple limiting value, $\gamma_{N/2}\to
\frac{1}{2}$.

\section{Discussion}

In summary, we investigated two first-passage problems involving
ordered random walks in one dimension. In both cases, there is a
series of survival probabilities and a nontrivial family of
first-passage exponents.  A universal function describes the exponents
when the number of particles is very large.  Remarkably, there are
scaling laws for the scaling exponents.

In general, a first-passage process with $N$ random walks is
equivalent to diffusion in a high-dimensional space with a complicated
absorbing boundary.  This boundary is typically formed by multiple
intersecting planes.  Solving the first-passage problem is equivalent
to solving the electrostatic problem given these boundary conditions
\cite{bk,cj,jdj,ntv}. For example, to find $\beta_1=0.228$ for a
three-particle system, we must solve for the potential in the exterior
of an insulating three-dimensional corner. Yet, the solution, which
likely requires an ingenious implementation of the image method,
remains unknown even in this physically relevant geometry \cite{dr}.

To circumvent this difficulty, we introduced the cone approximation in
which the absorbing boundary is replaced with the surface of a
suitably chosen cone.  This approximation is straightforward to
implement and utilizes exact and asymptotic properties for
first passage in a cone. The cone approximation provides useful
estimates for the exponents and moreover, it faithfully captures both
typical and extremal features. In particular, the cone approximation
yields the correct scaling variable for the first-passage exponents.

We used the term ``cone approximation'', yet this framework produces
the lower bound whenever the allowed space is a generalized cone with
the crucial property of invariance under dilation, ${\bf r}\to a {\bf
r}$ \cite{cone}. The computation of the first-passage exponent in a
generalized $d$-dimensional cone requires computation of the lowest
eigenvalue $\lambda$ of the angular portion $\nabla^2$ of the
Laplacian
\begin{equation}
\label{LB}
\nabla^2 f = -\lambda f, 
\end{equation}
subject to the Dirichlet boundary condition, $f=0$, on the surface of
the cone.  The choice of the smallest eigenvalue ensures that $f>0$
inside the allowed region. The first-passage exponent $\beta$ is
related to the lowest eigenvalue via $\lambda=2\beta(2\beta+d-2)$, see
\cite{bk}.

We now invoke a theorem that, in its simplest form, states that
amongst all domains with the same volume, the lowest eigenvalue of the
Laplace-Dirichlet operator occurs when the domain is a ball.  Rayleigh
conjectured this result for two dimensions \cite{jwsr}, and Faber and
Krahn eventually proved it \cite{ch}.  The Rayleigh-Faber-Krahn
theorem generalizes to higher dimensions and under mild conditions, to
Riemannian manifolds \cite{ic}. In the context of first-passage
processes, this theorem implies that of all generalized cones of fixed
solid angle, the first-passage exponent is minimal for the circular
cone. Since the regions discussed in our study qualify as generalized
cones, results of the cone approximation constitute rigorous lower
bounds for the exponents.

Cones can also provide upper bounds. To establish an upper bound, we
choose a cone that is inscribed by the absorbing boundary. Clearly,
the first-passage process is faster inside the inscribed cone and
therefore, the corresponding exponent must be an upper bound.

The cone approximation appears to be asymptotically exact for the
leadership problem as it predicts the bulk of the exponents with
excellent accuracy. A first step toward proving this exactness is the
observation that all opening angles in \eqref{alpha-lead} approach a
right angle as the number of particles diverges. The other major
challenge is finding an exact, or at least asymptotically exact,
framework for calculating the family of exponents in the
origin-crossing problem.

Finally, we mention that in the absence of a theoretical method for
obtaining statistics of first-passage exactly, Monte Carlo simulations
play vital role. We utilized a straightforward simulation
technique. It is interesting to find out how accelerated Monte Carlo
simulations fare in producing accurate estimates for the exponents
\cite{dba,pg,obdkgs}.

\medskip
This research has been supported by DOE grant DE-AC52-06NA25396 and
NSF grant CCF-0829541.


\begin{thebibliography}{99}

\bibitem{wf}
      W.~Feller,
      {\it An Introduction to Probability Theory and Its Applications}, vol. I
      (Wiley, New York, 1968).

\bibitem{ghw}
      G.~H.~Weiss,
      {\it Aspects and Applications of the Random Walk}
      (North-Holland, Amsterdam, 1994).

\bibitem{hcb}
       H.~C.~Berg,
       {\it Random Walks in Biology}
       (Princeton University Press, Princeton, 1983).

\bibitem{rg}
       J.~Rudnick and G.~Gaspari,
       {\it Elements of the Random Walk:
       An Introduction for Advanced Students and Researchers}
       (Cambridge University, Press, New York, 2004).

\bibitem{mef}
      M.~E.~Fisher,
      J.~Stat. Phys. {\bf 34}, 667 (1984).

\bibitem{hf}
      D.~A.~Huse and M.~E.~Fisher,
      Phys. Rev. B {\bf 29}, 239 (1984).

\bibitem{wbl}
     Q.~H.~Wei, C.~Bechinger, and P.~Leiderer,
     Science {\bf 287}, 625 (2000).

\bibitem{cdl}
     B.~X.~Cui, H.~Diamant, and B.~H.~Lin,
     Phys. Rev. Lett. {\bf 89}, 188302 (2002).

\bibitem{bpl}
     B.~P.~Lee,
     J. Phys. A {\bf 27}, 2633 (1994).

\bibitem{sr}
      S. Redner,
      {\it A Guide to First-Passage Processes}
      (Cambridge University Press, New York, 2001).

\bibitem{dhp}
      B.~Derrida, V.~Hakim, and V.~Pasquier,
      Phys. Rev. Lett. {\bf 75}, 751 (1995).

\bibitem{snm}
      S.~N.~Majumdar,
      Current Science {\bf 77}, 370 (1999).

\bibitem{hn}
      H.~Niederhausen,
      Eur. J. Combinatorics {\bf 4}, 161 (1983).

\bibitem{bdve}
       E.~Ben-Naim, Z.~A.~Daya, P.~Vorobieff, and R.~E.~Ecke,
       Phys. Rev. Lett. {\bf 86}, 1414 (2001).

\bibitem{fg}
      M.~E.~Fisher and M.~P.~Gelfand,
      J. Stat.\ Phys.\ {\bf 53}, 175 (1988).

\bibitem{bg} M. Bramson and D. Griffeath,
     in: {\it Random Walks, Brownian Motion, and Interacting
     Particle Systems: A Festshrift in Honor of Frank Spitzer}, eds.\
     R. Durrett and H. Kesten (Birkh\"auser, Boston, 1991).

\bibitem{djg}
       D.~J.~Grabiner,
       Ann. Inst. Poincare: Prob. Stat. {\bf 35}, 177 (1999).

\bibitem{kr}
       P.~L.~Krapivsky and S.~Redner, J.\ Phys.\ A {\bf 29}, 5347 (1996);
       S.~Redner and P.~L.~Krapivsky, Amer. J. Phys. {\bf 67}, 1277 (1999).

\bibitem{bjmkr}
      D.~ben-Avraham, B.~M.~Johnson, C.~A.~Monaco, P.~L.~Krapivsky,
      and S.~Redner,  J.~Phys.~A {\bf 36}, 1789 (2003).

\bibitem{ck}
      J.~Cardy and M.~Katori,  J.~Phys.~A {\bf 36}, 609 (2003).

\bibitem{dlb}
      D.~L.~Burkholder, Adv. Math. {\bf 26}, 182 (1977).

\bibitem{yal}
      S.~B.~Yuste, L.~Acedo, and K.~Lindenberg,
      Phys. Rev. E {\bf 64}, 052102 (2001).

\bibitem{gib}
      G.~I.~Barenblatt,
      {\it Scaling, Self-Similarity, and Intermediate Asymptotics}
      (Cambridge University Press, Cambridge, 1996).

\bibitem{hes}
       H.~E.~Stanley,
       {\it Introduction to Phase Transitions and Critical Phenomena}
       (Oxford University Press, New York, 1971).

\bibitem{bk}
       E.~Ben-Naim and P.~L.~Krapivsky,
       Kinetics of First Passage in a Cone, 
       preprint.

\bibitem{teh}
      T.~E.~Harris,
      J. Appl. Prob. {\bf 2}, 323 (1965).

\bibitem{dgl}
     D.~G.~Levitt,
     Phys. Rev. A {\bf 6}, 3050 (1973).

\bibitem{ap}
      S.~Alexander and P.~Pincus,
      Phys. Rev. B. {\bf 18}, 2011 (1978).

\bibitem{vkk}
      H.~van Beijeren, K.~W.~Kehr, and R.~Kutner,
      Phys. Rev. B {\bf 28}, 5711 (1983).

\bibitem{tw}
      C.~A.~Tracy and H.~Widom,
      Commun. Math. Phys. {\bf 159}, 151 (1994)

\bibitem{ra}
      R.~Arratia,
      Ann. Probab. {\bf 11}, 362 (1983).

\bibitem{krb}
      P. L. Krapivsky, S. Redner, and E. Ben-Naim,
      {\it  A Kinetic View of Statistical Physics}
      (Cambridge University Press, Cambridge, 2010).

\bibitem{rdd}
      R.~D.~DeBlassie,
      Probab. Theory Relat. Fields {\bf 74}, 1 (1987);
      Probab. Theory Relat. Fields {\bf 79}, 95 (1988).

\bibitem{bs}
      R.~Ba\~nuelos  and R.~G.~Smiths,
      Probab. Theory Relat. Fields {\bf 108}, 299 (1997).

\bibitem{NIST}
      {\em NIST Handbook of Mathematical Functions}, ed. F. W. J. Olver,
      D. M. Lozier,  et al. (Cambridge University Press, Cambridge, 2010).

\bibitem{largest} 
      To obtain the largest exponent $\beta_N$, we use
      $\alpha\to 1/2$ when $N\to\infty$. For opening
      angles $\alpha<\pi/2$, the survival exponent grows linearly with
      dimension: $\beta_N\simeq B\,N$ with
      $B^{\rm cone}=\frac{1}{4}\big[(\sin\tfrac{1}{2})^{-1}-1\big]$.

\bibitem{bh}
      E.~Ben-Naim and N.~W.~Hengartner,
      Phys.~Rev. E {\bf 76}, 026106 (2007).

\bibitem{bmr}
      D.~ben-Avraham, S.~N.~Majumdar, and S.~Redner,
      J. Stat. Mech. L04002 (2007).

\bibitem{abk}
      T.~Antal, E.~Ben-Naim, and P.~L.~Krapivsky, 
      J. Stat. Mech. P07009 (2010). 

\bibitem{cj}
      H.~S.~Carslaw and J.~C.~Jaeger,
      {\em Conduction of Heat in Solids}
      (Clarendon Press, Oxford, 1959).

\bibitem{jdj}
      J.~D.~Jackson,
      {\em Classical Electrodynamics}
      (Wiley, New York, 1998).

\bibitem{ntv}
      N.~Th.~Varopoulos,
      Math. Proc. Camb. Phi. Soc. {\bf 125}, 335 (1999);
      Math. Proc. Camb. Phi. Soc. {\bf 129}, 301 (1999).

\bibitem{dr}
      L.~C.~Davis and J.~Reitz,
      J. Math. Phys. {\bf 16}, 1219 (1975).

\bibitem{cone} 
      A generalized cone is an infinite convex domain that
      contains a special point, called an apex, and has the property
      that each ray emanating from the apex and going through any
      point inside a cone lies inside the cone.
     
\bibitem{jwsr}
     J.~W.~S.~Rayleigh, {\em The Theory of Sound}
     (Macmillan, New York, 1877; reprinted Dover, New York, 1945).

\bibitem{ch}
     R. Courant and D. Hilbert, {\em Methods of Mathematical Physics}, vol. I
     (Wiley, New York, 1953).

\bibitem{ic}
      I. Chavel, {\it Eigenvalues in Riemannian geometry} (Academic
      Press, Orlando, 1984).

\bibitem{dba} D.~ben-Avraham, J. Chem. Phys. {\bf 88}, 941 (1988).

\bibitem{pg}
      P.~Grassberger,
      Computer Phys. Comm. {\bf 147}, 64 (2002).

\bibitem{obdkgs}
      T.~Oppelstrup, V.~V.~Bulatov, A.~Donev, M.~H.~Kalos,
      G.~H.~Gilmer, and B.~Sadigh,
      Phys. Rev. E {\bf 80}, 066701 (2009).


\end{thebibliography}
\end{document}